\documentclass[nocompress]{spie}  

\usepackage{amsmath,amsfonts,amssymb}
\usepackage{graphicx}
\usepackage[colorlinks=true, allcolors=blue]{hyperref}

\title{Performance characterization and near-realtime monitoring of MUSE adaptive optics modes at Paranal}

\author[1]{T. Wevers}
\author[1]{F. Selman}
\author[1,2]{A. Reyes}
\author[1,3]{M. Vega}
\author[1,4]{J. Hartke}
\author[1]{F. Bian}
\author[6,5]{O. Beltramo-Martin}
\author[5,7]{R. JL. F\'etick}
\author[8]{S. Kamann}
\author[9]{J. Kolb}
\author[1]{T. Kravtsov}
\author[10]{C. Moya}
\author[5]{B. Neichel}
\author[9]{S. Oberti}
\author[11]{C. Reyes}
\author[9]{E. Valenti}

\affil[1]{European Southern Observatory, Alonso de C\'ordova 3107, Vitacura, Santiago, Chile}
\affil[2]{Universidad Técnica Federico Santa María, Avenida España 1680, Valparaíso, Chile}
\affil[3]{Instituto de Astronom\'ia, Universidad Cat\'olica del Norte,
Angamos 0610, Antofagasta, Chile}
\affil[4]{Sub-Department of Astrophysics, University of Oxford, Denys Wilkinson Building, Keble Road, Oxford OX1 3RH, United Kingdom}
\affil[5]{Aix Marseille Univ, CNRS, CNES, LAM, Marseille, France}
\affil[6]{SpaceAble, Paris, France}
\affil[7]{ONERA, 29 avenue de la Division Leclerc, 92322 Chatillon, France}

\affil[8]{Astrophysics Research Institute, Liverpool John Moores University, Liverpool, United Kingdom}
\affil[9]{European Southern Observatory, Garching bei Munchen, Germany}
\affil[10]{ Instituto de Astrofísica, Facultad de Física, Pontificia Universidad Católica de Chile, Santiago, Chile.}
\affil[11]{School of Physics, University of New South Wales, Sydney, NSW 2052, Australia}
\authorinfo{Further author information: (Send correspondence to T.W., F.S. or F.B.)\\T.W.: E-mail: twevers@eso.org\\  F.S.: E-mail: fselman@eso.org\\ F.B.: E-mail: fbian@eso.org}

\begin{document} 
\maketitle

\begin{abstract}
The Multi Unit Spectroscopic Explorer (MUSE) is an integral field spectrograph on the Very Large Telescope Unit Telescope 4, capable of laser guide star assisted and tomographic adaptive optics using the GALACSI module. Its observing capabilities include a wide field (1 square arcmin), ground layer AO mode (WFM-AO) and a narrow field (7.5"$\times$7.5"), laser tomography AO mode (NFM-AO). The latter has had several upgrades in the 4 years since commissioning, including an optimisation of the control matrices for the AO system and a new sub-electron noise detector for its infra-red low order wavefront sensor. We set out to quantify the NFM-AO system performance by analysing $\sim$230 spectrophotometric standard star observations taken over the last 3 years. To this end we expand upon previous work, designed to facilitate analysis of the WFM-AO system performance. We briefly describe the framework that will provide a user friendly, semi-automated way for system performance monitoring during science operations. We provide the results of our performance analysis, chiefly through the measured Strehl ratio and full width at half maximum (FWHM) of the core of the point spread function (PSF) using two PSF models, and correlations with atmospheric conditions. These results will feed into a range of applications, including providing a more accurate prediction of the system performance as implemented in the exposure time calculator, and the associated optimization of the scientific output for a given set of limiting atmospheric conditions.
\end{abstract}


\section{INTRODUCTION}
\label{sec:intro}  
The depth and image quality of (optical and infra-red) observations taken at ground-based observatories are at the mercy of the prevailing atmospheric conditions. In particular, atmospheric turbulence will set the width and shape of the long exposure point spread function (PSF). As a result, the full-width at half maximum (FWHM) of the PSF is typically much larger than the theoretical diffraction limit achievable by a primary mirror of diameter $D$ at wavelength $\lambda$: FWHM $>>$ $\frac{\lambda}{D}$. 

The introduction of an adaptive optics system can drastically change this picture. By using natural and/or laser guide stars (NGS / LGS), it is possible to measure the (average) distortions that a wavefront incurs by passing through the Earth's atmosphere. These distortions can be (partially) removed by applying corrections with a deformable mirror in the optical train. Such corrections flatten the wavefront prior to arriving at the science instrument. Techniques using LGSs are being developed and deployed at a rapid pace, and will play a crucial role in the instrumentation of the next generation of extremely large telescopes (see e.g. \cite{maory}). They have a large range of applications, from providing a diffraction limited spatial resolution over improved astrometric measurements to allowing observations to reach higher signal to noise ratios, maximising the scientific return of deep field observations.

The Multi Unit Spectroscopic Explorer (MUSE\cite{2010SPIE.7735E..08B}) instrument is a 24-channel integral field spectrograph mounted in the Nasmyth B focus of the Very Large Telescope (VLT)'s unit telescope 4 (UT4, named Yepun in the native Mapudungun language). Between the telescope and MUSE there is a module, the Ground Atmospheric Layer Adaptive Corrector for Spectroscopic Imaging, GALACSI\cite{2012SPIE.8447E..37S}, that passes a corrected wavefront to MUSE. GALACSI employs a set of 4 sodium LGSs operating at a wavelength of 594~nm (the 4 Laser Guide Star Facility, or 4LGSF\cite{2014SPIE.9148E..3OH}) to feed its dedicated AO system consisting of four 40x40 Shack-Hartmann wavefront sensors (WFSs). Corrections are computed via a real-time system, SPARTA (Standard Platform for Adaptive
optics Real Time Applications\cite{2006SPIE.6272E..10F}), and subsequently sent to the 1170 voice-coils actuators of the deformable secondary mirror (DSM). 

MUSE has two AO modes: one wide field mode (WFM) with a field of view (FoV) covering 1 square arcmin, and a narrow field mode (NFM) covering 7.5" by 7.5". For WFM, the LGS constellation is positioned on the 4 sides of the wide FoV, and the AO system is conjugated to sense the ground layer contribution (i.e the lower $\sim$800 m) of the turbulence -- this is dubbed ground layer AO, or GLAO. While the AO system senses the higher order modes of turbulence, the low order modes (tip and tilt) are sensed with a dedicated (off axis) tip/tilt star (TTS) outside of the MUSE FoV. For NFM, the LGSs are positioned close to the science object of interest. The use of a (on axis) NGS in combination with the LGSs allows for a full reconstruction of the turbulence profile in the atmosphere, and is therefore called laser tomography AO (LTAO). 

Continual improvement of various components of the AO system is desirable as new techniques and better detectors become available. Quantifying the performance of an AO system is therefore ideally a dynamic task, both to identify potential problems and to optimize the scientific return, and to provide the user community with the latest available performance estimates to plan future projects. Arguably the biggest challenge for quantifying AO performance is the need for dedicated on-sky observations. This is especially problematic for instruments such as MUSE, with an oversubscription rate of requested versus available time of roughly 10\footnote{\url{https://www.eso.org/sci/publications/announcements/sciann17445.html}}. Another challenge is the multidimensional nature of the problem. The AO system performance depends on atmopsheric factors such as coherence time, seeing, and airmass, but also wind-speed, full turbulence profile, telescope vibrations, etc.

In this contribution we expand upon the work described in Hartke et al.\cite{Hartke20}, which presents the performance of the MUSE WFM-AO mode. Here we focus on performance verification and characterization of the NFM-AO mode of MUSE. We set out to provide this verification using spectrophotometric standard star (hereafter STD) observations, rather than using dedicated on-sky observations, in order to minimize the impact on regular science operations. These STD observations are part of the calibration plan, and are therefore taken every night that science data is taken in NFM-AO. This ensures the availability of adequate data, and provides the opportunity to incorporate the NFM-AO performance measurements into regular operation routines. We note that this is only possible because the calibrations for NFM-AO are taken in full AO mode (i.e. the AO system is providing real-time corrections); for WFM-AO this is not the case: standard star observations for the WFM-AO modes are taken without LGSs. The other advantage of using standard star fields is that source detection becomes trivial, as these observations contain only a single, bright object. We have devised our measuring process to allow, in the future, for the use of regular scientific observations of stellar fields to monitor AO performance. This will significantly increase the number of observations available for performance monitoring.

Before we delve into the performance characterization of the NFM-AO mode, we first report on recent upgrades of the IR low order sensor (IRLOS) used to sense the tip, tilt, defocus and astigmatism modes with the NGS. We further extend the database framework presented in Hartke et al. (2020) to include all AO modes of MUSE. While currently we only incorporate basic environmental and performance metrics, this approach provides a scalable and flexible environment to include further AO and telescope telemetry. In the future this may prove useful for analysing and further optimising the AO system performance as a whole.

\section{CHANGES AND IMPROVEMENTS TO MUSE NFM SINCE COMMISSIONING}

\begin{table*}[ht!]
    \label{tab:nfm_log}
    \begin{tabular}{lp{14cm}}
    \hline
        \textbf{Date} & \textbf{Event description} \\
    \hline
    \hline
    2018 May 7 & First commissioning run of NFM is completed. \vspace{3mm}\\
    
2018 Sept 5 -- 18 & Science verification observations are done for NFM. \vspace{3mm}\\

2018 Oct 1 & NFM starts regular science operations. \vspace{3mm}\\
    
2019 Nov 18 & New control matrices for NFM have been installed providing an improved performance which in some cases can be a factor of 2.  \vspace{3mm}\\
  
2021 March 25 & IRLOS upgraded with a new SAPHIRA sub-electron read-out mode detector. Requirements for TT stars are now: \newline (1) for non-extended reference sources the system can go down to J=16.5 in general conditions and J=17.0 for turbulence category (TC) 10\%, both now driven at 500Hz instead of 200Hz. \newline (2) For extended sources the limiting magnitude is J=14.5, driven at 200Hz.  \vspace{3mm}\\
    
2021 July 18 & A new faint mode was implemented and offered using the SH wavefront sensor at 200 Hz. This allows the use of NGSs as faint as magnitude J=18.5 for all TCs, and down to J=19.0 for TC=10\% and airmasses below 1.2. \\
    \hline
    \end{tabular}
\caption{Important events in the life of the MUSE NFM-AO mode.}
\end{table*}

\subsection{AN IMPROVED TOMOGRAPHY RECONSTRUCTION}
The commissioned control architecture of GALACSI NFM is based on a single Matrix Vector Multiply (MVM) Minimum Mean Square Error (MMSE) reconstructor \cite{Oberti2018}. An upgrade of the LTAO reconstruction algorithm was later implemented and verified at the end of 2019. Because of the LGS Tip/Tilt indetermination, the reconstruction of the defocus and astigmatism modes is no longer done in the altitude layers (i.e. these are only reconstructed for the ground layer). As a result, the regularization (i.e. the imposing of priors within the algorithm to counteract the ill-posedness of the reconstruction problem) could be fine-tuned with a more aggressive gain along the higher spatial frequencies, and in turn to allow for the correction of a larger number of modes ($\sim$850). The temporal controller was also fined tuned to improve the temporal error. In addition, the redefinition of the valid subaperture maps and tweaks to the noise and phase covariance matrices were implemented. Finally, analysis of the error budget revealed a dominant contribution related to vibrations, most likely associated with a mechanical origin in another instrument which was subsequently removed. 

The main result of these changes is an improvement of the Strehl ratio performance by a factor of $\sim$2 (at 940 nm). In addition, more modes can be controlled by the LTAO system, which allows MUSE NFM-AO to be operated under a wider range of (i.e. worse) atmospheric conditions. Visible PSFs recorded with the commissioning camera sampled at 5 mas per pixel confirmed the increase of angular resolution with full width at half maximum as low as 28 milliarcsec.

\subsection{AN UPGRADE OF THE WAVEFRONT SENSOR: IRLOS+}
MUSE NFM-AO provides science data at optical wavelengths (480--930~nm). Because LTAO requires the NGS to be nearly on-axis, in practice the science target is often also used as the NGS. For optimal performance the low order turbulence modes are therefore sensed at IR rather than at optical wavelengths. This provides the additional advantages of a smaller PSF (as FWHM $\propto \lambda^{-6/5}$), smaller errors in the WFS centroid measurements as well as a higher SNR, facilitating a better correction (the limiting factor is typically the number of available photons, especially for faint sources). 

The upgraded IRLOS detector (dubbed IRLOS+) uses a SAPHIRA Mark14b detector with sub-electron readout noise\cite{2019SPIE11180E..6LF}. During commissioning this was shown to add at least two magnitudes to the limit for the wavefront sensing while simultaneously increasing the correction rate to 500 Hz. The 500 Hz small scale mode (compared to the old small scale mode operated at 200 Hz) of the IRLOS+ system is in operations with a fully automated acquisition sequence since the beginning of Period 107, with a limiting magnitude of J=16.0 for point sources. This fainter magnitude limit was offered for P108 and proposals submitted during P107\footnote{Prior to this commissioning MUSE NFM-AO could use only AO reference stars brighter than H=14.5.}. A faint mode is offered since the second commissioning, which ended 2021 July 18. With the faint mode the limiting magnitude has been extended by approximately 4 mags down to J=19.0.
During the IRLOS+ commissioning we were able to close the loop and get reasonable corrections using stars as faint as J=19.3 (which translates to 2.1 photons per frame per sub-aperture). 

The modes supported by IRLOS+ are detailed in Table~\ref{tab:irlosplus_modes}.  Prior to IRLOS+ there was a single small scale and a single large scale mode clocked at 200~Hz, with a limiting magnitude of approximately H=14 in small scale. For small scale modes, the IRLOS+ FoV is divided into 20x20 pixels (with a pixel scale of 78 mas per pixel), representing a size of 1.56x1.56 arcsec. For large scale modes, the pixel scale is 314 mas per pixel, translating into a FoV of 6.28x6.28 arcsec. A much more versatile suite of modes is now available for different NGS magnitudes for both point and extended sources.

\begin{table*}[h!]
\label{tab:irlosplus_modes}
\begin{tabular}{lccrrp{4cm}}
\hline
J-band & Name & 4LGSF & Gain & Freq. & Filter \\
    mag        & AOS MAIN MODES IRLOS           &   &   & Hz  &  \\
\hline\hline
\multicolumn{6}{c}{Small scale: pix: 78 mas, FoV 20 pix x 20 pix – 1.56” x 1.56”} \\
\hline
0.5..3.5   & 20x20\_SmallScale\_500Hz\_LowGain  & W & 1 & 496 & 1600BW20nm \\
3.5..5.5   & 20x20\_SmallScale\_500Hz\_LowGain  & W & 1 & 496 & 1600BW190nm \\
5.5..7.0   & 20x20\_SmallScale\_500Hz\_LowGain  & W & 1 & 496 & CLEAR \\
7.0..10.5  & 20x20\_SmallScale\_500Hz\_LowGain  & N & 1 & 496 & CLEAR \\
10.5..16.0 & 20x20\_SmallScale\_500Hz\_HighGain & N & 68 & 496 & CLEAR \\
16.0..19.0 & 20x20\_SmallScale\_200Hz\_HighGain & N & 100 & 200 & CLEAR\_8as or CLEAR \\
\hline
\multicolumn{6}{c}{Large scale: pix: 314 mas, FoV 20 pix x 20 pix -- 6.28” x 6.28”} \\
\hline
7..10.5 & 20x20\_LargeScale\_200Hz\_LowGain & N & 1 & 200 & CLEAR \\
10.5..17 & 20x20\_LargeScale\_200Hz\_HighGain & N & 68 & 200 & CLEAR \\
\hline
\end{tabular}

\caption{Description of the IRLOS+ new operating modes. J-band gives the magnitude range of the TT star to select the mode described in the row. Column Name gives the name of the mode as it appears in the headers with key ``AOS MAIN MODES IRLOS''. Column 4LGSF indicate the size of the laser constellation; N for the nominal size, and W for the wide constellation that avoids contaminating the LGS wavefront sensors (at some cost in terms of Strehl). The column Gain gives the gain setup for the SAPHIRA detector. Column Freq. gives the frequency of the tip-tilt correction. And finally, column Filter indicates the filter used for IRLOS; CLEAR\_8as is a filter mask combination that prevents star images from one lenslet to be visible in other quadrants. It is used when acquiring the TT star in crowded fields.}
\end{table*}

\newpage
\section{DATA, PROCESSING AND PRODUCTS}
\label{sec:data}  

\begin{figure*}[h!]
    \includegraphics[width=\textwidth, keepaspectratio, angle=0,origin=c]{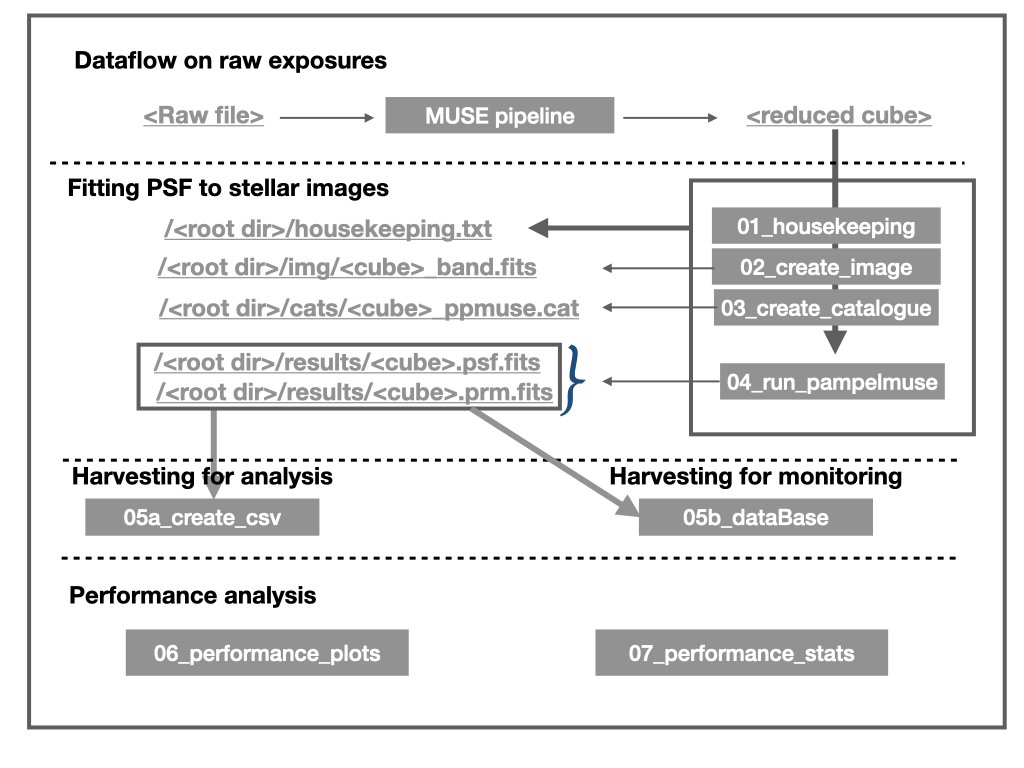}
    \caption{Schematic representation of the data flow designed for AO performance analysis and monitoring. The dashed lines separate parts of the data-flow handled by independent, asynchronous, processes.}
    \label{fig:flowchart}
\end{figure*}

\subsection{OBSERVATIONS AND DATA PROCESSING}
We collect all NFM-AO STD observations taken after 2019 November 18, following the optimization of the control matrices that are incorporated in the LTAO algorithm. This provides a dataset of 229 observations. The data processing can be divided into 3 broad steps, and closely follows the procedures described in detail in Hartke et al. 2020. We briefly reiterate the processing steps, and highlight where the procedure is different for NFM-AO compared to WFM-AO:

\begin{enumerate}
    \item {\bf Data flow on raw exposures}\\ We retrieve the data, reduced by the MUSE data reduction pipeline\cite{2020A&A...641A..28W}, from the science archive. When working in near real-time the reduced data will be automatically processed by the pipeline within the UT4 dataflow system. 
    
    \item {\bf Fitting PSF to stellar images} 
    \begin{enumerate}
    \item We create images in the Cousins I-band and perform source detection with the SExtractor package. Unlike scientific data, the STD observations currently used for NFM-AO do not include world coordinate system information in the frame headers. We therefore work in pixel space. The source detection list should contain a single bright object.
    \item To perform the PSF fitting, we use two software packages: (1) {\tt muse\_nfm\_psflib} which was developed in house for the initial NFM commissioning, and then extended for the IRLOS+ commissioning; and (2)  the PampelMuse software\cite{2013A&A...549A..71K, 2018ascl.soft05021K} coupled to the Modelization of the Adaptive Optics PSF in PYthon ({\tt maoppy}) PSF model\cite{2019A&A...628A..99F, Fetick20}. {\tt muse\_nfm\_psflib} is used to perform PSF fits using a dual Moffat profile and the {\tt maoppy} model in custom made scripts. Within PampelMuse, we use the SINGLESRC routine rather than the INITFIT routine (the former is the equivalent of the latter but works in pixel space instead of sky coordinates) to determine the PSF star. We provide an initial guess for the FWHM of 0.1 arcsec and use a PSF radius of 2.5 pixels. We leave all the {\tt maoppy} parameters free to vary, allowing for translations as well as asymmetries (i.e. ellipticity) in the PSF. The advantage of PampelMuse over the in-house script is that it permits the analysis of crowded stellar fields, therefore it is better suited for the near real-time analysis of science observations containing stellar fields.
    \item The output of these steps is a file, the PSF file, for each observation containing all relevant PSF parameters and images of the data, the model, and the residuals.
    \end{enumerate}
    \item {\bf Harvesting for analysis} \\
    For the offline analysis of data from the archive, which is what we mostly present in this contribution, we extract from the PSF files, either from our in-house scripts or from PampelMuse, performance parameters and write them in tables in csv files. For near real-time work at the telescope we have created a database script that reads from the PSF files all keywords and performance parameters from binary extensions, and then write it into a MariaDb database.
    \item {\bf Performance analysis} \\
    The data is analyzed via python scripts and/or Jupyter notebooks that either read directly from the database or from the csv files. A Python class has been defined for the near real-time analysis and it will be presented elsewhere.
\end{enumerate}

A flow chart representing these steps is shown in Figure \ref{fig:flowchart}. The intention is to make the PSF files, and the database and the csv files available to outside users.

\subsection{PERFORMANCE CHARACTERIZATION}
\begin{figure*}[h!]
    \centering
    \includegraphics[width=\textwidth]{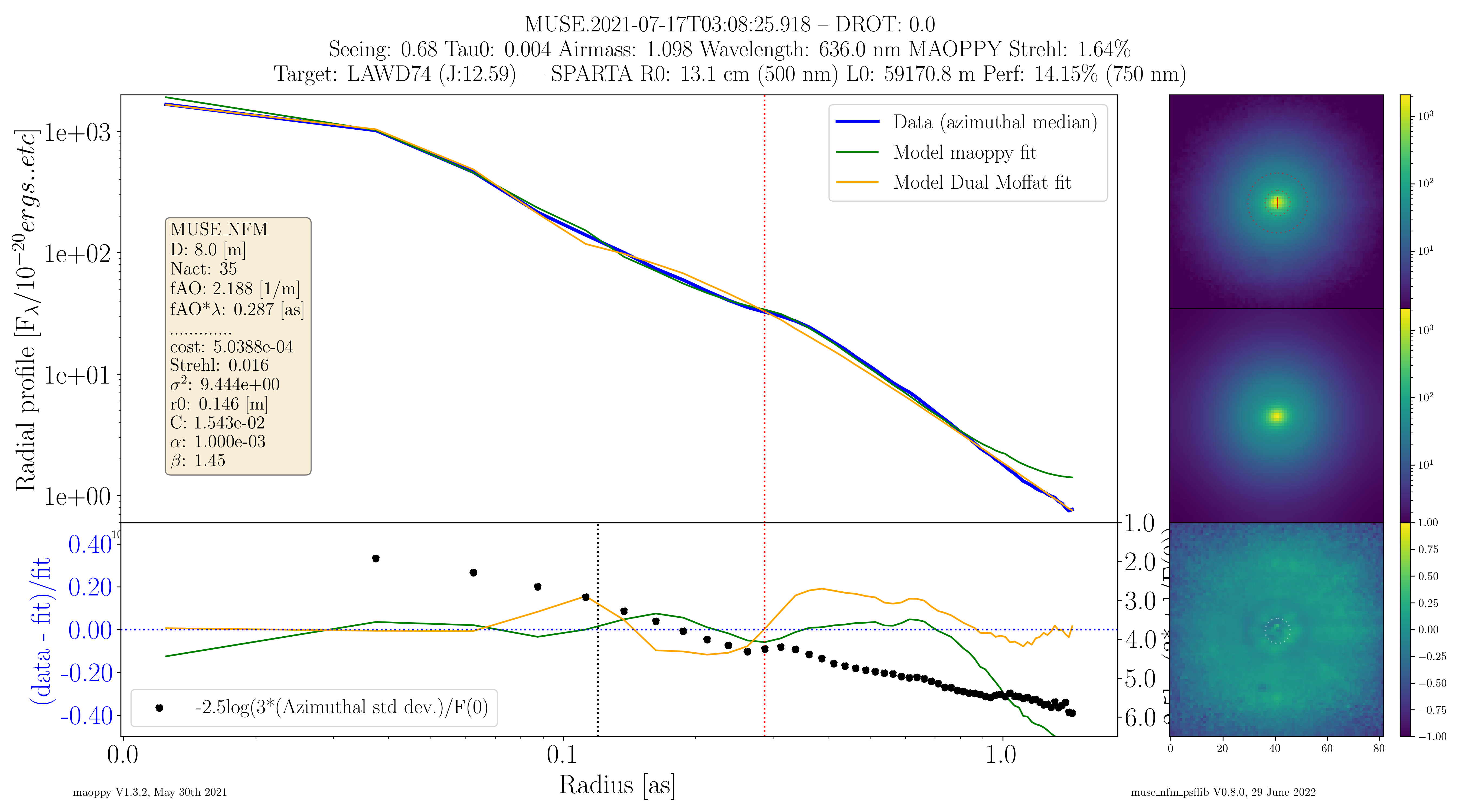}
    \caption{Examples of PSF fits using a double 2D Moffat and a {\tt maoppy} PSF. The right side column with image stamps shows the data, the fit, and the residuals of the {\tt maoppy} fit from top to bottom, respectively. For reference we overplot circles at 0.120 arcsecs and at the maximum-correction radius for the wavelength displayed, 0.287~arcsec at 636~nm. The upper-left panel shows the profiles extracted from the 2D panels: data in blue, {\tt maoppy} fit in green, and Moffat fit in orange. The bottom-left panel shows the residuals, and an estimate of the contrast.}
    \label{fig:fit_residuals}
\end{figure*}

The performance of an LTAO system can be characterised in various ways. In this work we adopt two traditional metrics to quantify the level of performance: the FWHM of the PSF within the AO correction radius (i.e. the FWHM of the narrow PSF core) and the Strehl ratio of the PSF.
The Strehl ratio is defined as the ratio of the peak of the aberrated PSF to the theoretical (diffraction limited) PSF. We must caution that for NFM-AO, the path through the MUSE optics introduces aberrations which can be similar to, or even larger than, the FWHM of the diffraction limited PSF coming from the telescope. Furthermore, the MUSE detectors undersample the PSF and therefore the measured Strehl is highly biased, as shown by previous work.

To explore the application of model fitting for predictive purposes (for example, in the exposure time calculator) we use two models to provide a parametrized description of the PSF: i) a dual Moffat fit to the PSF (Section \ref{sec:moffat}), and ii) a model with a physical basis, the {\tt maoppy} model as implemented by the {\tt maoppy} software version 1.3.2 (Section \ref{sec:maoppy}). The {\tt maoppy} PSF (as well as a detailed example involving MUSE NFM data) is described in detail in \cite{2019A&A...628A..99F, Fetick20}.
We will quantify and cross-validate the model fitting results with the values obtained directly from the data in the next section.

For reference, the design requirement was for MUSE+GALACSI to provide a Strehl ratio of 5 per cent at 650 nm in NFM-AO.
As a baseline, in the following we provide all parameters at a wavelength of $\sim$650 nm. This wavelength is chosen because it is the reference wavelength of the NFM-AO mode, and many of the quantities measured by the AO system (see Section \ref{sec:environment}) are provided at this value. We also explore the behaviour of the variables with wavelength. Where relevant we will also investigate the observed behaviour at the extreme blue and red wavelengths probed by the NFM-AO system (500 nm and 900 nm, respectively).\\

\subsubsection{PERFORMANCE VARIABLES: DUAL MOFFAT PSF MODEL}
\label{sec:moffat}
 The first model is a poor person's approach and assumes that the total PSF can be described by a double Moffat profile. Two Moffat functions are fit simultaneously to the data, with free FWHM and beta parameters. The expectation was that a broad Moffat will describe the seeing-limited wings of the PSF (corresponding to the region outside the AO correction radius), and a narrow Moffat function will describe the core of the PSF (representing the AO corrected region). The advantage of this model is that it is highly flexible, easy to use, and familiar to most astronomers non-specialist in AO. Furthermore, its parameters have an unambiguous physical meaning, making their interpretation straight forward. The drawback is that it does not take into account the underlying physics. For example, the AO corrected PSF likely has a more complex shape depending on the system performance, including a non-monotonically decreasing intensity profile with radius that cannot be captured in the Moffat profile. The main parameters of interest in this model are the FWHM of the inner (narrow) Moffat profile (fit to the PSF core) and the associated Strehl ratio. We find in general, and as can be seen in Figure~\ref{fig:fit_residuals}, that the wide Moffat is not as wide as the uncorrected seeing halo, and the narrow component is narrower than the core of the PSF. This is perhaps the reason why the Moffat estimates of the Strehl ratio are always larger than the {\tt maoppy} ones, but it should be pointed out that the analysis of this is work in progress. Note that the radial plots in Figure~\ref{fig:fit_residuals} are determined as azimuthal medians from the 2D frames.

\subsubsection{PERFORMANCE VARIABLES: MAOPPY PSF MODEL}
\label{sec:maoppy}
The second approach utilises the {\tt maoppy} PSF model, which employs a more sophisticated underlying physical model of the AO-corrected and uncorrected regions. It can be shown using basic Fourier analysis that an AO system cannot correct aberrations with a spatial wavelength shorter than the inter-actuator distance, giving rise to a maximum-correction radius (in image space). The core of the PSF is determined by the main-mirror diameter and the quality of the corrections to the incoming wavefront. The fit is done using the inverse of the variances as weights.

Figure \ref{fig:fit_residuals} illustrates an example of PSF fits to a standard star observation, as well as a plot of the ratio of the residuals over the fits (to provide an estimate of the relative goodness of fit). The right column of  Figure~\ref{fig:fit_residuals} shows 82x82 pixels stamps, centered on the reference star, corresponding to approximately 2x2 arcseconds: the actual data, top; the fitted {\tt maoppy} model, center; and the residuals (i.e. data - model) bottom.

The main parameters of the {\tt maoppy} PSF fit are shown in the insert table in Figure \ref{fig:fit_residuals}. As shown in the figure, this model provides a reasonably accurate representation of the AO corrected PSF. We can see that the {\tt maoppy} residuals are significantly smaller than for the dual Moffat model, particularly around the maximum correction radius. It should be pointed out that Figure~\ref{fig:fit_residuals} corresponds to a very bright star which makes fitting errors more evident. One drawback is that most of the model parameters are more difficult to interpret directly. For this reason we extract from the fitted {\tt maoppy} 2D models a core FWHM, and a Strehl ratio (as provided by {\tt maoppy}'s otf\_strehl function).

\subsubsection{ATMOSPHERIC (ENVIRONMENT) VARIABLES}
\label{sec:environment}
During science operations, decisions for the execution of NFM-AO observations are made based on the atmospheric parameters measured by the Differential Image Motion Monitor (DIMM). 
The DIMM provides measurements of the seeing (corrected to zenith) at 500 nm, as well as a measurement of the coherence time ($\tau_0$). These measurements are performed using bright stars that are not along the line of sight, so can provide only rough approximations of the true conditions at the sky position of the observation. An additional available measurement is the image quality of the guide probe, which provides a FWHM measurement along the line of sight of the observation using the telescope guide star. This measurement is taken in a 100 nm wide band centred on 650 nm. Finally, the AO real time computer SPARTA also measures several quantities, including the seeing (at 500 nm), coherence time, Strehl ratio and Fried parameter (at 650 nm) based on data from the 4 LGSs. An important caveat for the SPARTA measurements is that the LGS photons do not traverse the MUSE optical path. These measurements will therefore likely provide optimistic values, compared to the measurements on the MUSE data (e.g. for the Strehl ratio).

For consistency, reported values in this work are provided around 650 nm. An exception is the analysis related to the turbulence categories (TCs), which are defined with the values measured at 500 nm by the DIMM. When appropriate, we perform corrections for seeing and coherence time values measured at different wavelengths assuming that seeing $\propto \lambda^{-1/5}$ and $\tau_0 \propto \lambda^{6/5}$. Where indicated, we account for airmass degradation of the seeing (coherence time) as $\propto$ 3/5 (--3/5) power, respectively.

\section{PRELIMINARY RESULTS}
\subsection{FRIED PARAMETER r$_0$}
\begin{figure}
    \centering
    \includegraphics[width=0.5\textwidth]{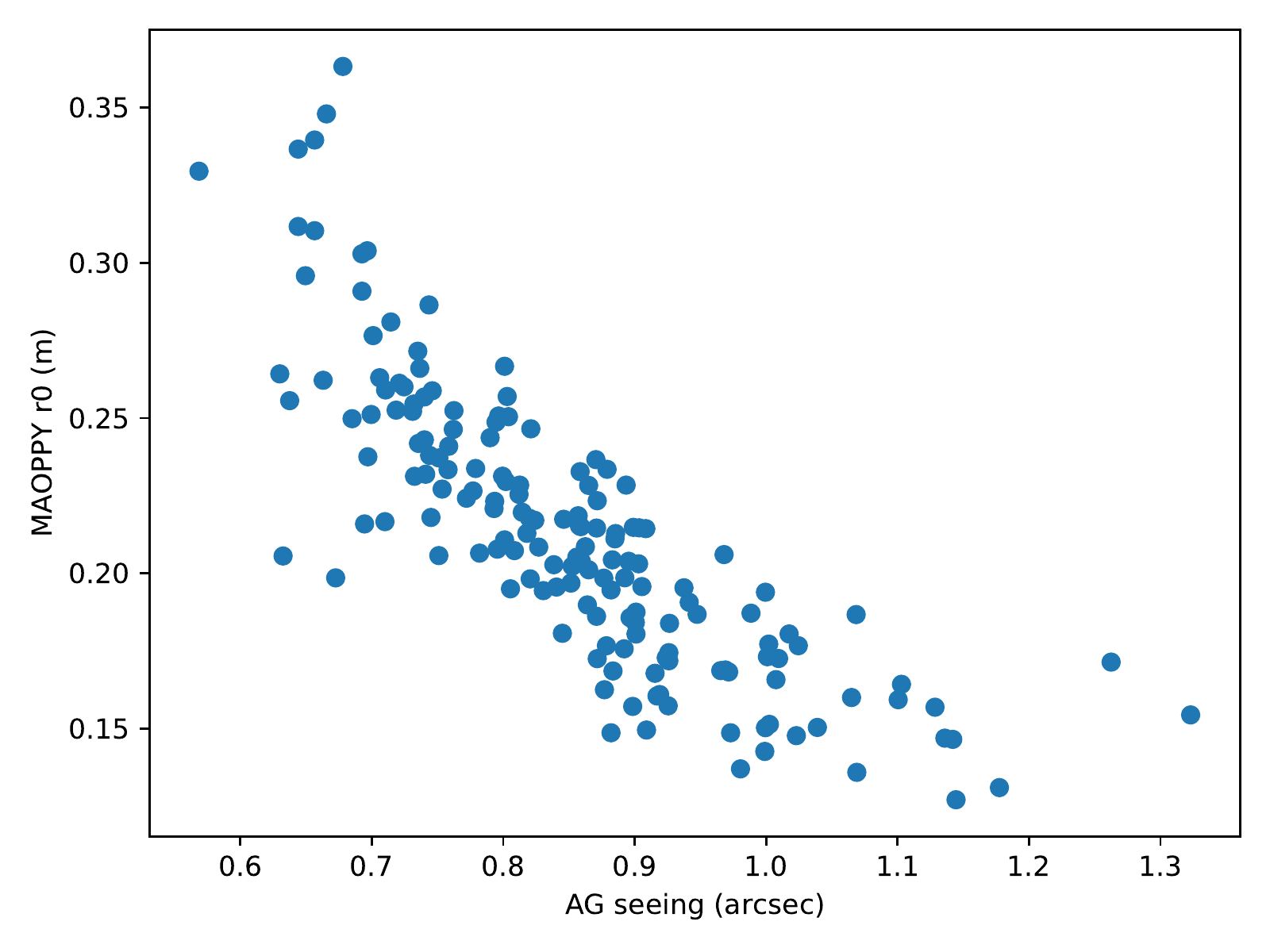}
    \caption{Fried parameter r$_0$, determined through PSF fitting with {\tt maoppy}, as a function of the line of sight (autoguider) seeing. A negative correlation is observed, as expected.}
    \label{fig:fried_seeing}
\end{figure}
\begin{figure}
    \centering
    \includegraphics[width=0.7\textwidth]{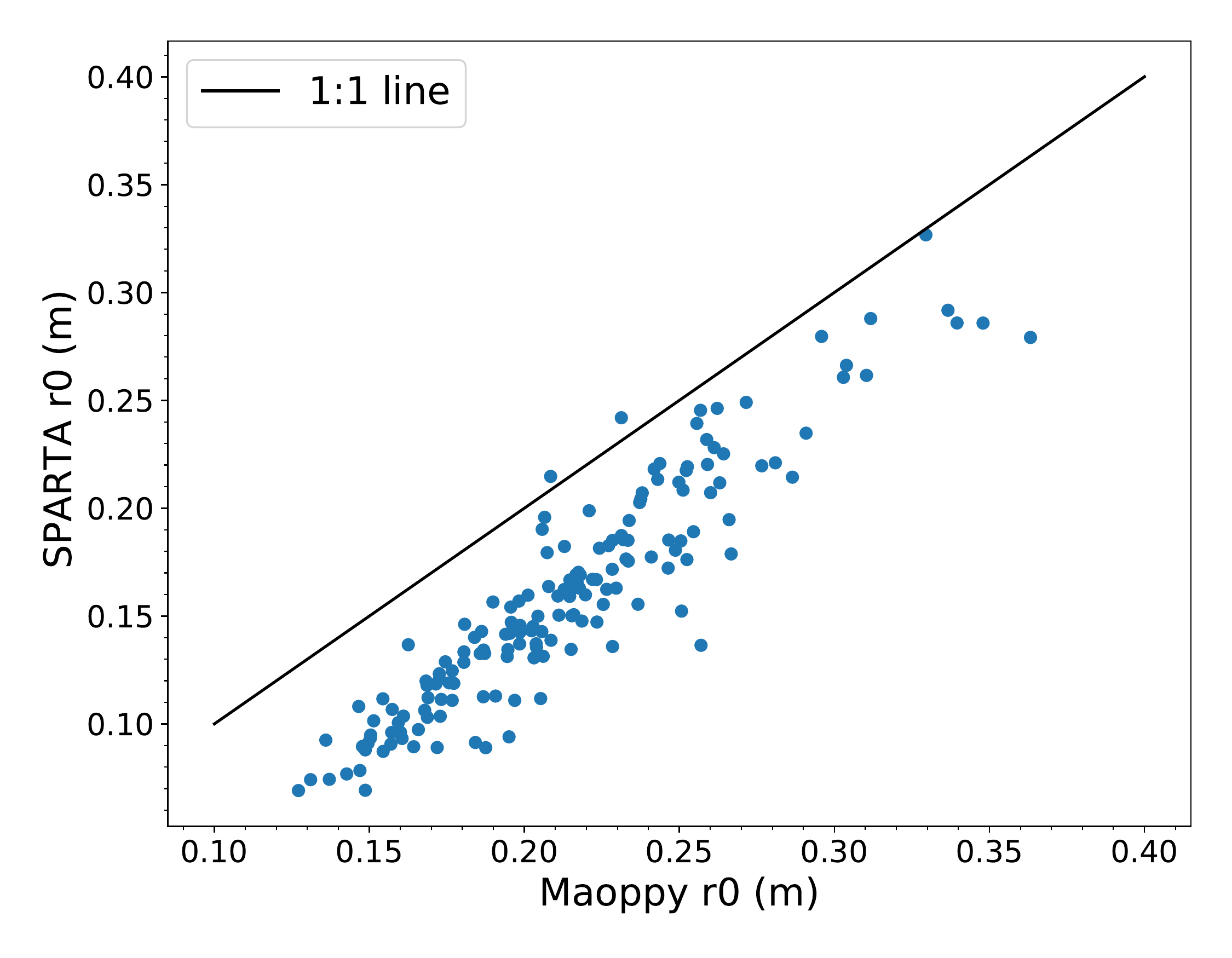}
    \caption{Comparison of the Fried parameter r$_0$ as measured by the SPARTA system and as determined from the {\tt maoppy} fitting. A small offset ($\sim$ 5 cm) is observed.}
    \label{fig:fried_sparta}
\end{figure}

Before we delve into the details of the NFM-AO system performance, we first verify that the fitting results follow the expectations in relation to turbulence properties. In Figure \ref{fig:fried_seeing} we show the {\tt maoppy} r$_0$ parameter as a function of the autoguider FWHM, a proxy for the line of sight seeing. A negative correlation is observed, as expected: if the atmosphere is more turbulent, then the Fried parameter becomes smaller. 
Figure \ref{fig:fried_sparta} shows the measured r$_0$ parameter through the SPARTA system with that inferred from the {\tt maoppy} fitting. The straight line indicates perfect agreement. There is an offset of $\sim$0.05 m, where {\tt maoppy} gives systematically larger values than SPARTA. As we will see later, this may help to explain the somewhat pessimistic estimates from {\tt maoppy} for e.g. the Strehl ratio, compared to the Moffat fitting results.\\

In the following sections we present the results of the performance metrics, as measured through the two models, as a function of environment parameters. We note that the fitting routines were not successful for all the available data. We therefore only present results based on aggregated data and focus on the observed trends, rather than going into a detailed analysis of the different models for given data points.

\subsection{PERFORMANCE AS A FUNCTION OF TURBULENCE CATEGORIES}
\begin{figure}
    \centering
    \includegraphics[width=0.48\textwidth]{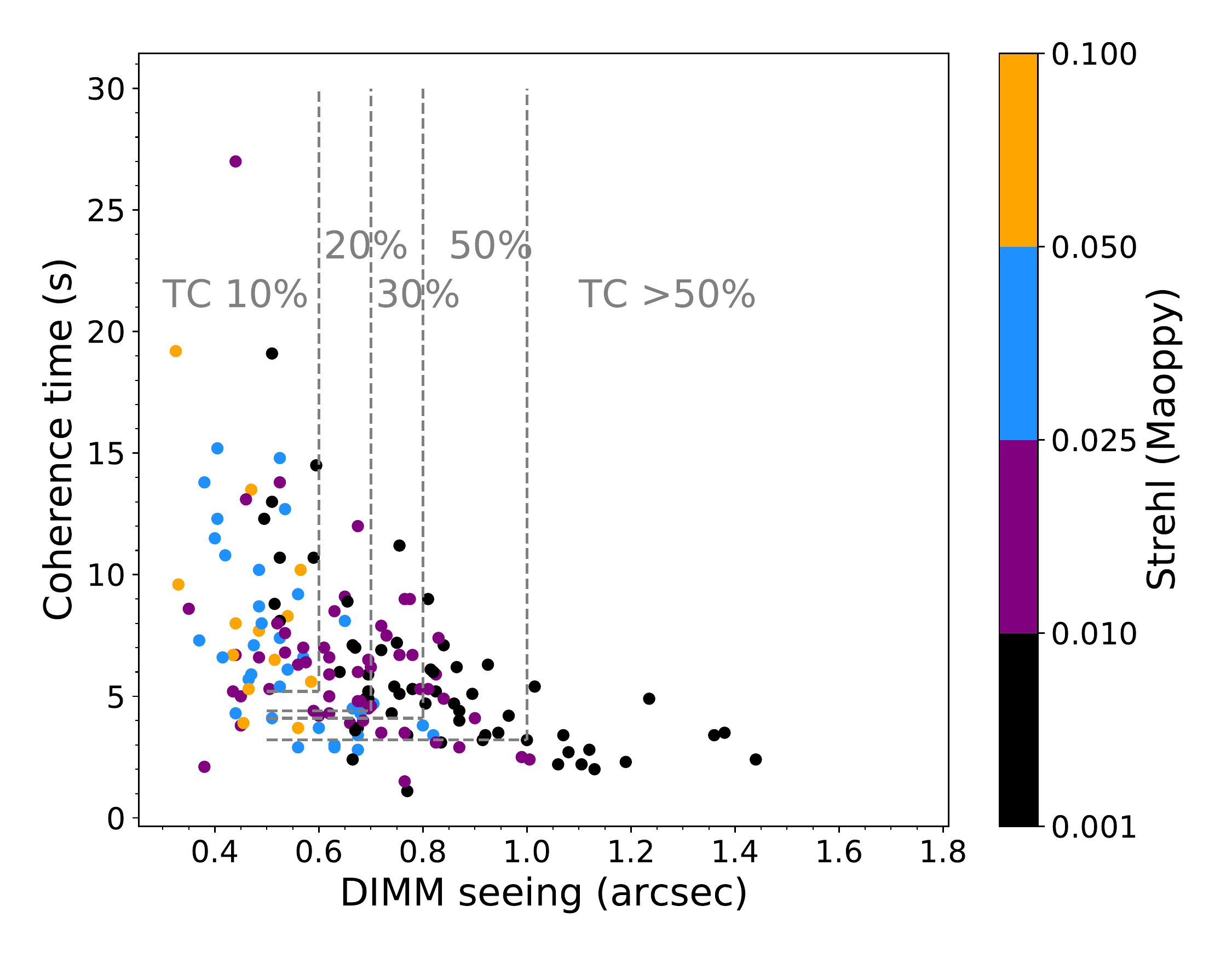}
    \includegraphics[width=0.48\textwidth]{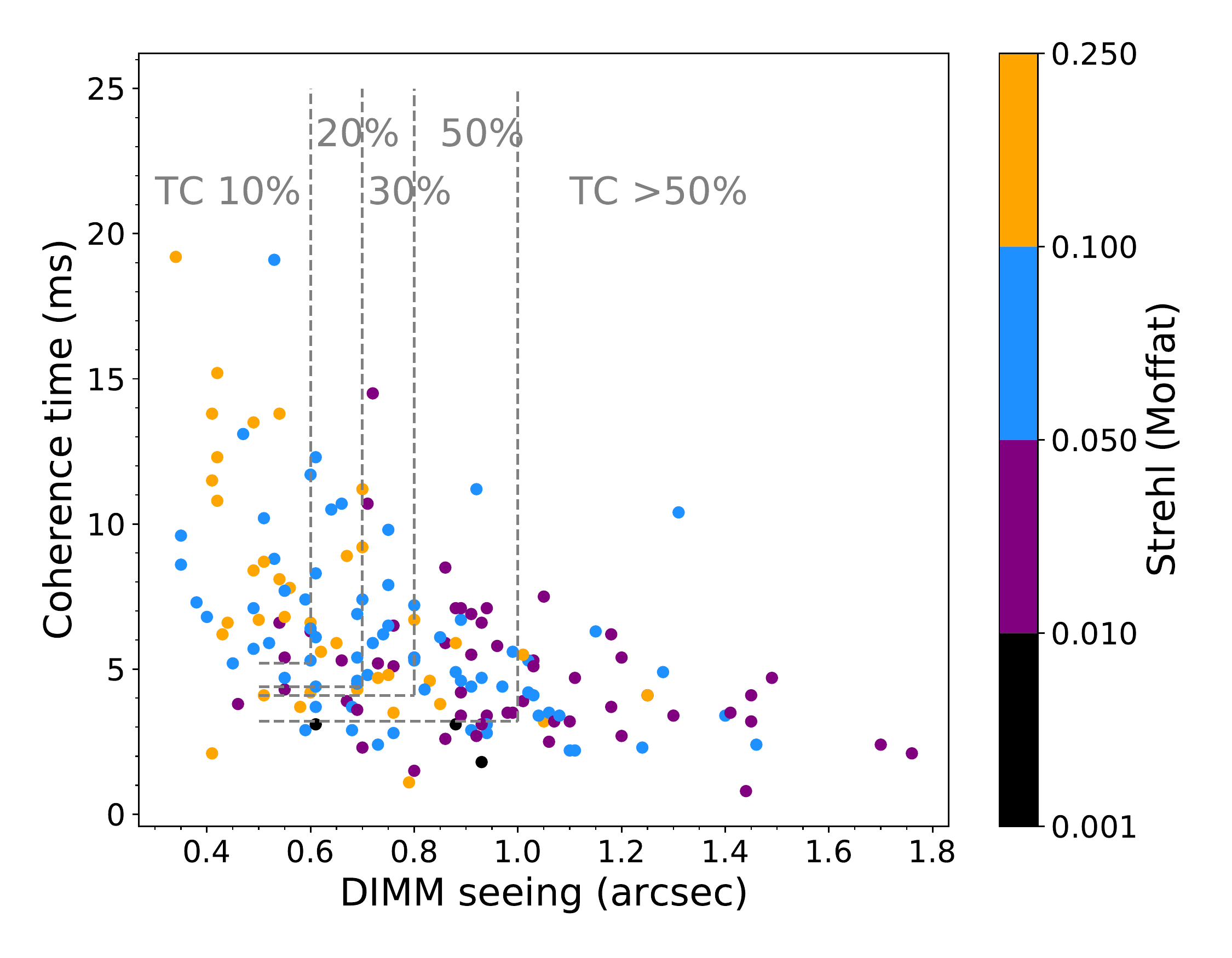}
    \caption{Performance of the NFM-AO system quantified according to the assigned turbulence category. From left to right (and top to bottom), the dashed lines indicate progressively worse turbulence conditions. TC10\% represents the best 10\% of observed conditions. The colour coding represents the Strehl ratio measured from the {\tt maoppy} (left panel) and Moffat (right panel) results, respectively. Note the difference in colour bar scales. There is a clear trend of improving Strehl ratio as a function of turbulence category.}
    \label{fig:TCcats_strehl}
\end{figure}

\begin{figure}
    \centering
    \includegraphics[width=0.48\textwidth]{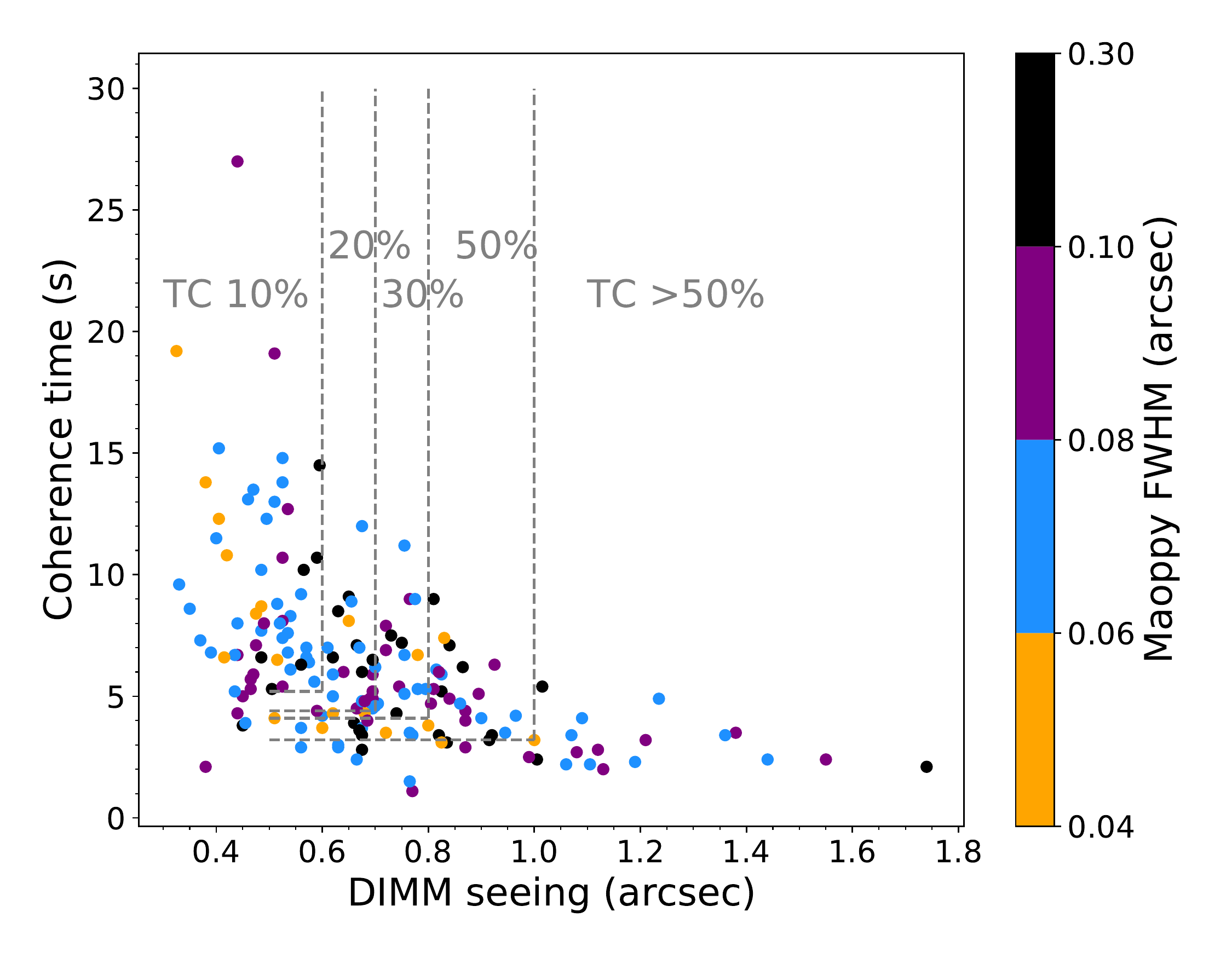}
    \includegraphics[width=0.48\textwidth]{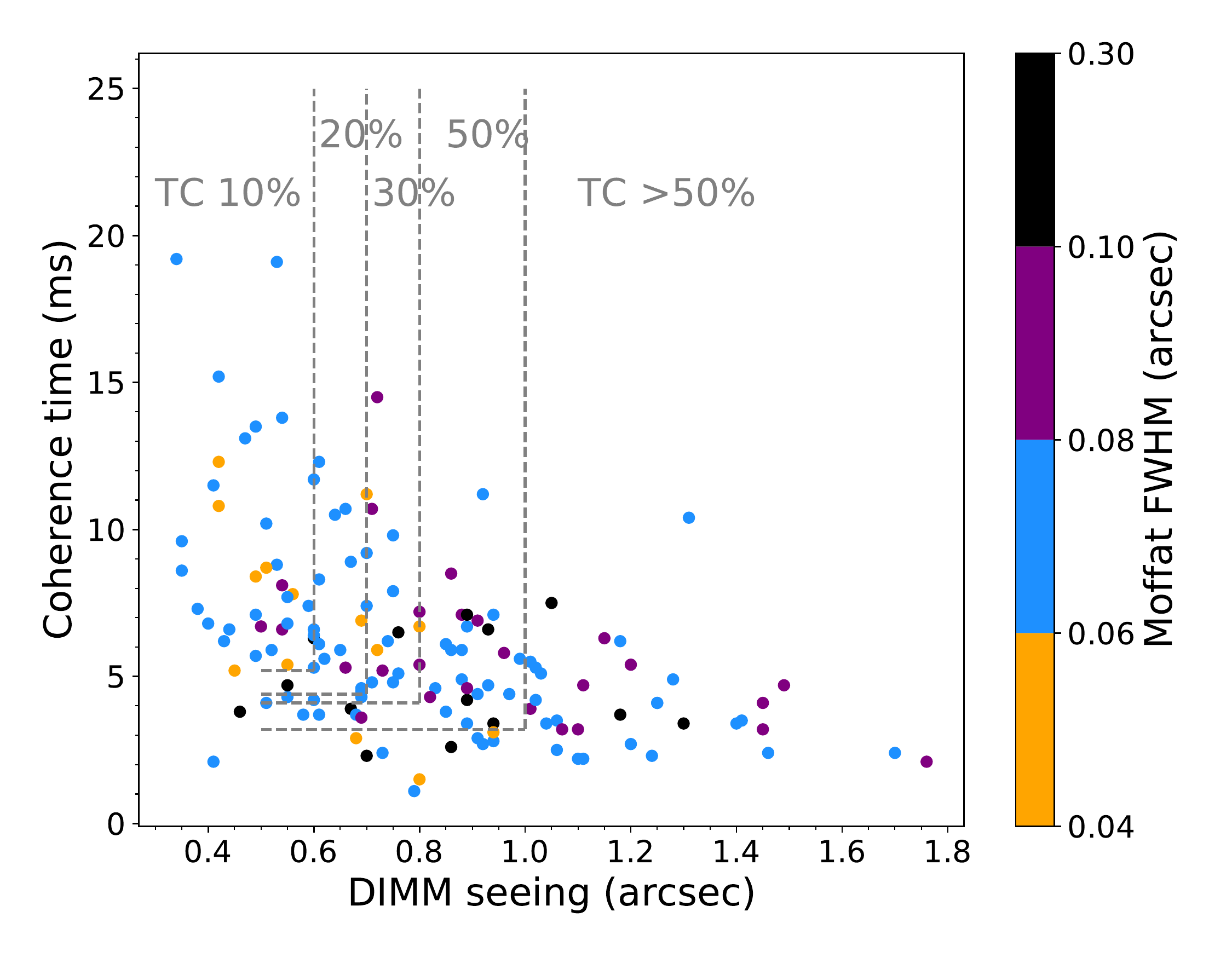}
    \caption{Same as Figure \ref{fig:TCcats_strehl}, but now with a colour coding according to the inner Moffat or {\tt maoppy} PSF FWHM. There is only a weak dependence of the FWHM on the turbulence category.}
    \label{fig:TCcats_fwhm}
\end{figure}
We present the performance of the NFM-AO system as a function of the defined TCs in Figures \ref{fig:TCcats_strehl} and \ref{fig:TCcats_fwhm}, showing the measured Strehl ratio and PSF FWHM obtained with the {\tt maoppy} and Moffat models, respectively. The various turbulence categories are represented by the dashed grey lines. The top left parameter space represents the best atmospheric conditions, and moving to the right and bottom directions leads to progressively worse TCs. These TCs represent the corresponding deciles in DIMM seeing and coherence time conditions at Paranal, and are defined as:
\begin{equation*}
\begin{array}{l}
   \rm TC10\%:  seeing < 0.6"\ and\ \tau_0 > 5.2\ \rm ms\\
  \rm  TC20\%:  seeing < 0.7"\ and\ \tau_0 > 4.4\ \rm ms\\
   \rm TC30\%: seeing < 0.8"\ and\ \tau_0 > 4.1\ \rm ms\\
   \rm TC50\%:  seeing < 1.0"\ and\ \tau_0 > 3.2\ \rm ms\\
    \end{array}
\end{equation*}

The minimum requirement for using NFM-AO during science operations is TC50\% or better, although the need to obtain spectrophotometric standard star observations (which are often observed in twilight to minimize pressure on night time by the calibration plan) sometimes requires that worse conditions are used. 

From inspection of Figure \ref{fig:TCcats_strehl}, the general trend is as expected: better atmospheric conditions result in higher Strehl ratios for both models. The effect of seeing appears to be more important than the effect of coherence time; this is explored in more detail in the next Section. The intrinsic scatter within each TC can be partially explained by the effects of the airmass of the observation, which will degrade the resulting Strehl ratio. We note that there is a significant difference between the {\tt maoppy} and Moffat Strehl ratios in terms of absolute values, with the former being smaller.
The Moffat results (right panel) indicate that a Strehl ratio of $>1$\% is reached in nearly all cases, with values up to $\sim$20\%. We hence conclude that MUSE is capable of achieving the design requirement (Strehl ratio of 5\%) in the majority of observations (caveats are discussed in more detail in Section \ref{sec:summary}). We further note that for the data points with a Strehl ratio $<5$\% (the purple and black points), the majority (44/57) were observed at airmass $>1.1$ (and 34 of these at airmass $>$1.2), highlighting the importance of airmass to the Strehl performance of NFM-AO.
Median performance metrics for each TC (which approaches 5 \% for TC50\%), which provide a rough estimate of the performance that can be expected, are reported in Table \ref{tab:performance_TC}. 
\begin{table}[]
    \centering
    \begin{tabular}{ccccc}
       TC  & $\lambda$ & FWHM & Strehl & \# \\
       & (nm) & (mas) & (per cent)\\ \hline
        10 & 498 & 81 (36) & 4.5 (2.5) & 35 \\
        & 590 & 71 (18) & 7.4 (3.9) \\
        & 698 & 69 (15) & 10.0 (4.1) & \\
        & 774 & 68 (15) & 13.7 (5.9) \\
        & 912 & 64 (11) & 19.3 (5.3)& \\
        20 & 498 & 91 (22) & 3.5 (1.6) & 22 \\
        & 590 & 78 (59) & 5.5 (2.5) \\
        & 698 & 70  (18) & 8.4 (2.2) & \\
        & 774 & 69 (11) & 12.3 (3.3) \\
        & 912 & 69 (22) & 16.4 (3.8)& \\
        30 & 498 & 88 (28) & 3.0 (2.1)& 21 \\
        & 590 & 76 (19) & 5.6 (3.2) \\
        &698 & 73  (25) & 7.9  (4.4)& \\
        & 774 & 74 (34) & 10.8 (5.0) \\
        & 912 & 68 (27) & 15.5 (4.9)& \\
        50 & 498 & 126 (63) & 1.4 (1.2)& 40 \\
        & 590 & 94 (36) & 2.8 (2.4) \\
        & 698 & 81 (34)  & 4.8 (3.6)&  \\
        & 774 & 75 (56) & 7.8 (4.0) \\
        & 912 & 71 (19) & 10.5 (4.6)& \\\hline
    \end{tabular}
    \caption{Estimates of the median performance for each TC (based on the Moffat PSF fitting), as well as the number of datapoints used to derive these results. Values between brackets denote the sample standard deviation per bin, providing an indication of the spread. The effect of airmass has not been accounted for.}
    \label{tab:performance_TC}
\end{table}

In addition to the Strehl ratio, we also defined the PSF FWHM as a performance metric. These results are shown in Figure \ref{fig:TCcats_fwhm}. Here our findings for both models are more consistent with each other; we find that a FWHM of 80 mas or better can be achieved in most observing conditions. 
The delivered FWHM does not show a strong trend with TC, although again we note a strong dependence on observation airmass, where 29/40 observations with FWHM$>$80 mas were taken at airmass $>1.1$ (and 24 of these have airmass $>1.2$). By measuring these quantities in different wavelength planes (as indicated in Table \ref{tab:performance_TC}) we can also investigate the behaviour of the FWHM with wavelength. This is expected to improve with wavelength as the AO system performs better for the redder wavelengths. While we find an improvement of $\sim$20--30\% when moving from the blue to the red part of the wavelength range, the relative difference is most pronounced at bluer wavelengths specially under bad conditions (as evidenced by the large improvement in going from 498 nm to 682 nm but the more modest improvement when moving to 912 nm). 

\subsubsection{PERFORMANCE AS A FUNCTION OF TURBULENCE CATEGORIES AND AIRMASS}
\begin{figure*}
    \centering
    \includegraphics[width=0.48\textwidth]{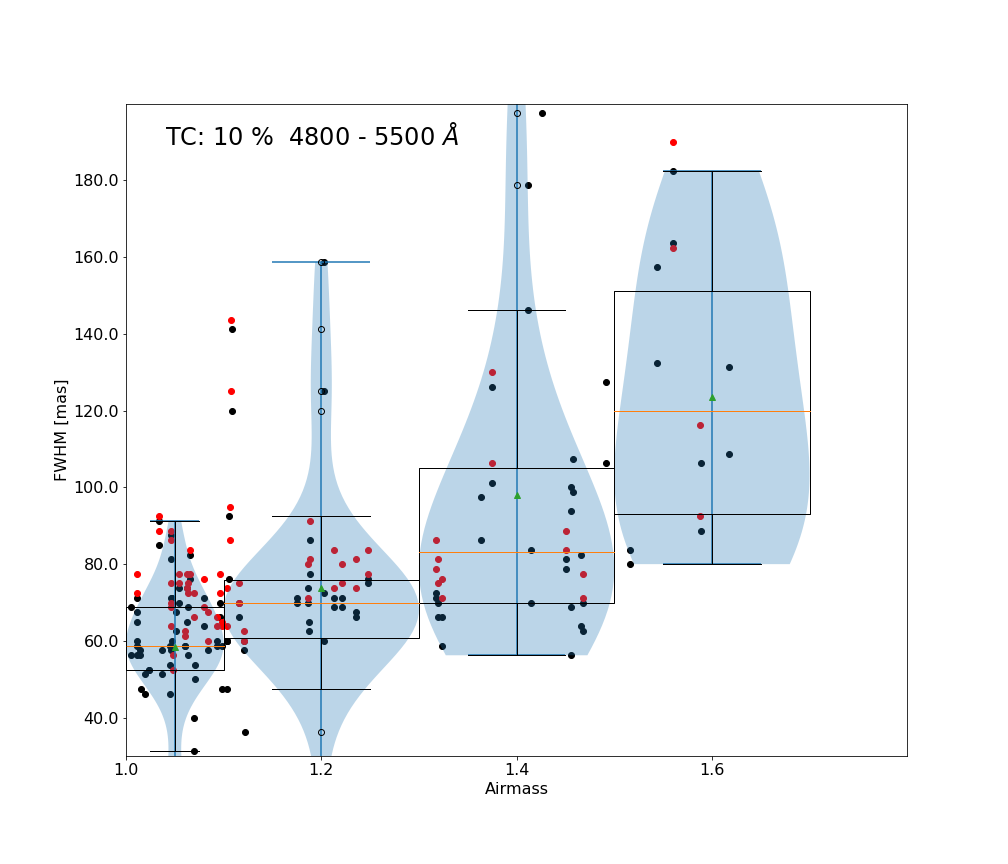}
    \includegraphics[width=0.48\textwidth]{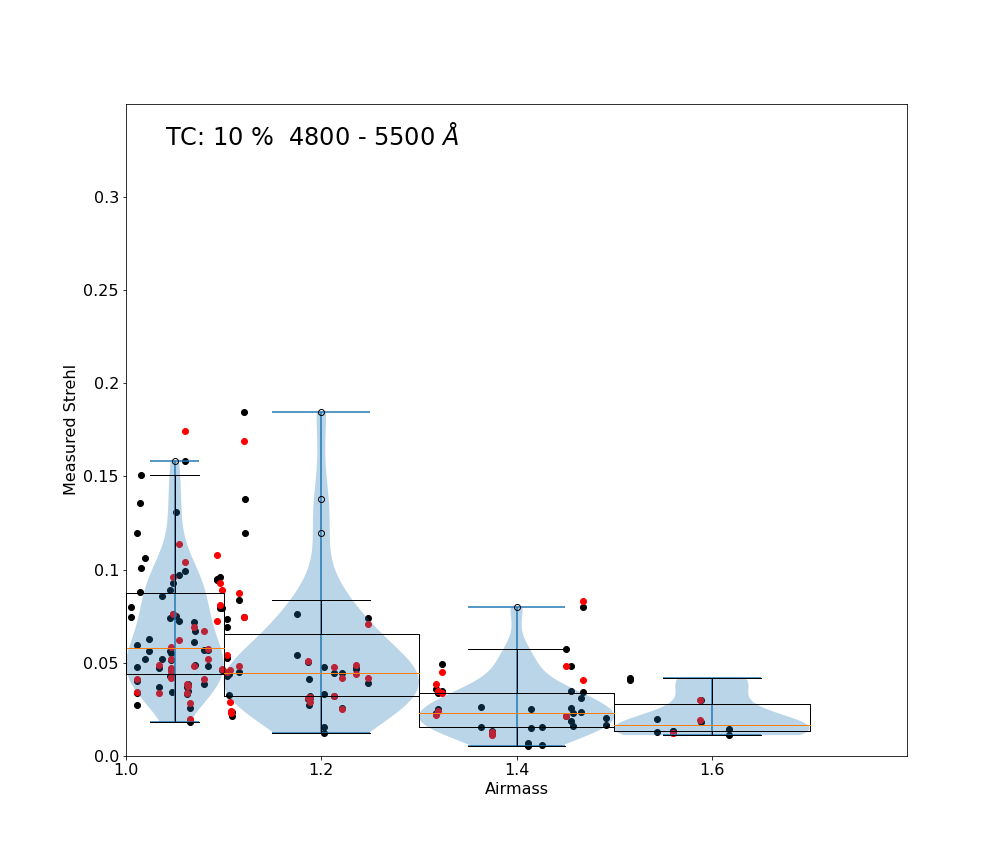}
    \includegraphics[width=0.48\textwidth]{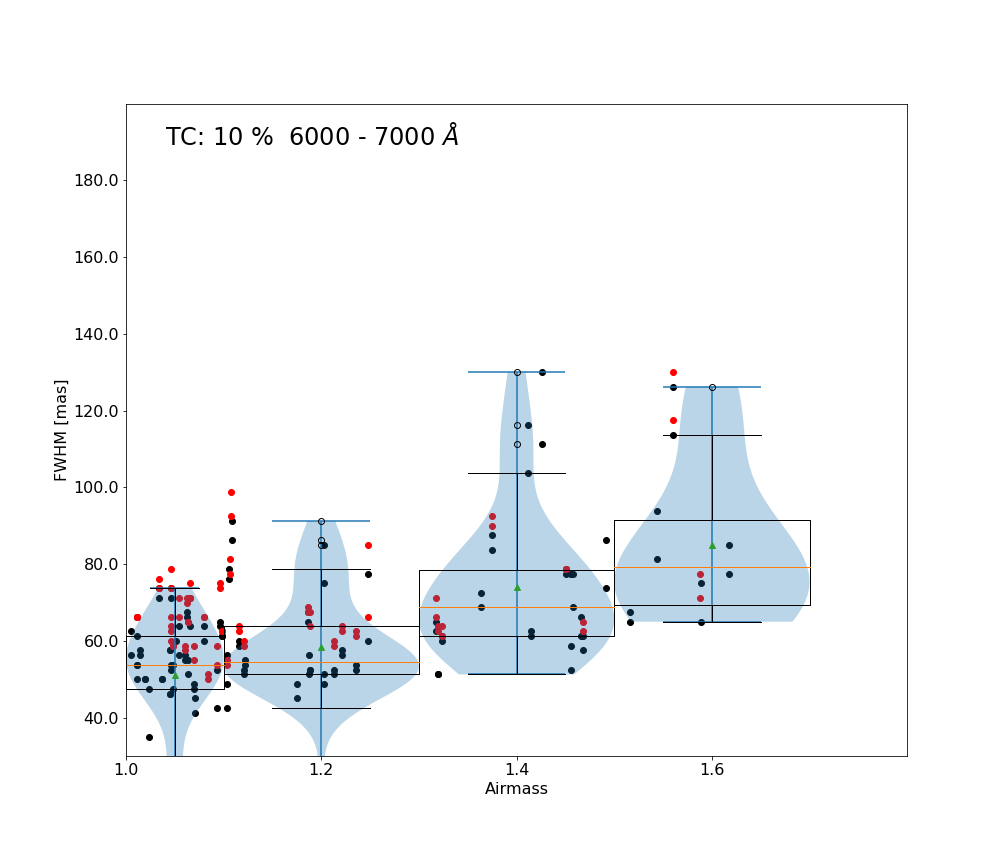}
    \includegraphics[width=0.48\textwidth]{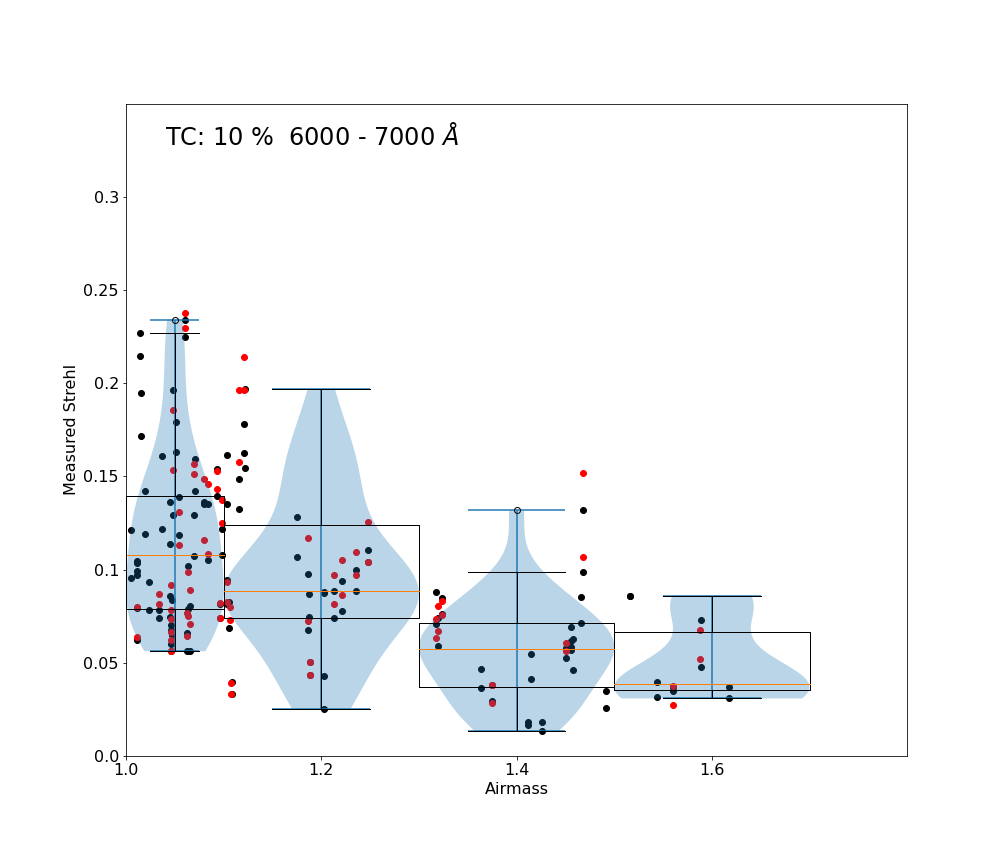}
    \caption{The FWHM and Strehl for TC 10\% observations at two different wavelength ranges as a function of airmass. Red points indicate data before the control matrix optimization.}
    \label{fig:Performance_vs_airmass}
\end{figure*}

We can disentangle the airmass effect from other variables by studying the performance, as measured by for example the PSF core FWHM or Strehl, versus airmass for the different turbulence categories. This is an important exercise, because currently users of the instrument specify the turbulence category and the maximum airmass of their observations in the ETC to determine the exposure time that will give them a given SNR. The current version of ESO's ETC uses a fixed NFM PSF that does not depend on TC or airmass. Preliminary results are presented in Figure~\ref{fig:Performance_vs_airmass}. This figure shows the deterioration of performance with airmass, which is particularly pronounced for the bluer wavelengths. 

\subsection{CORRELATION ANALYSIS}
\begin{figure}
    \centering
    \includegraphics[width=\textwidth]{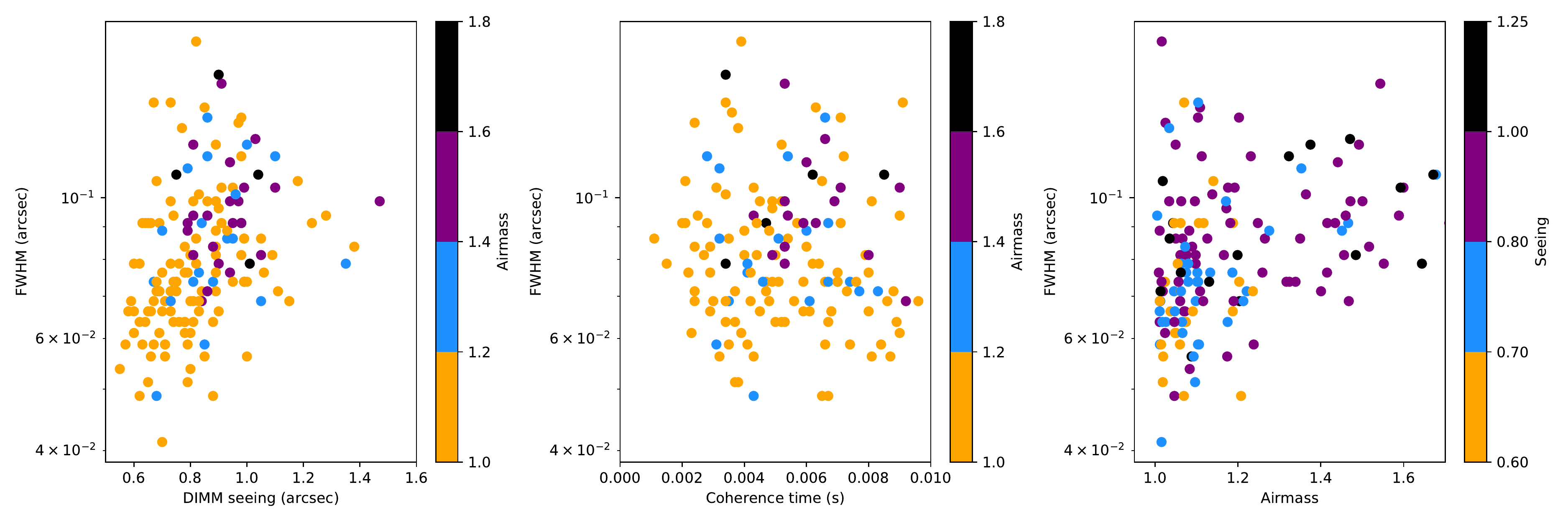}
    \centering
    \includegraphics[width=\textwidth]{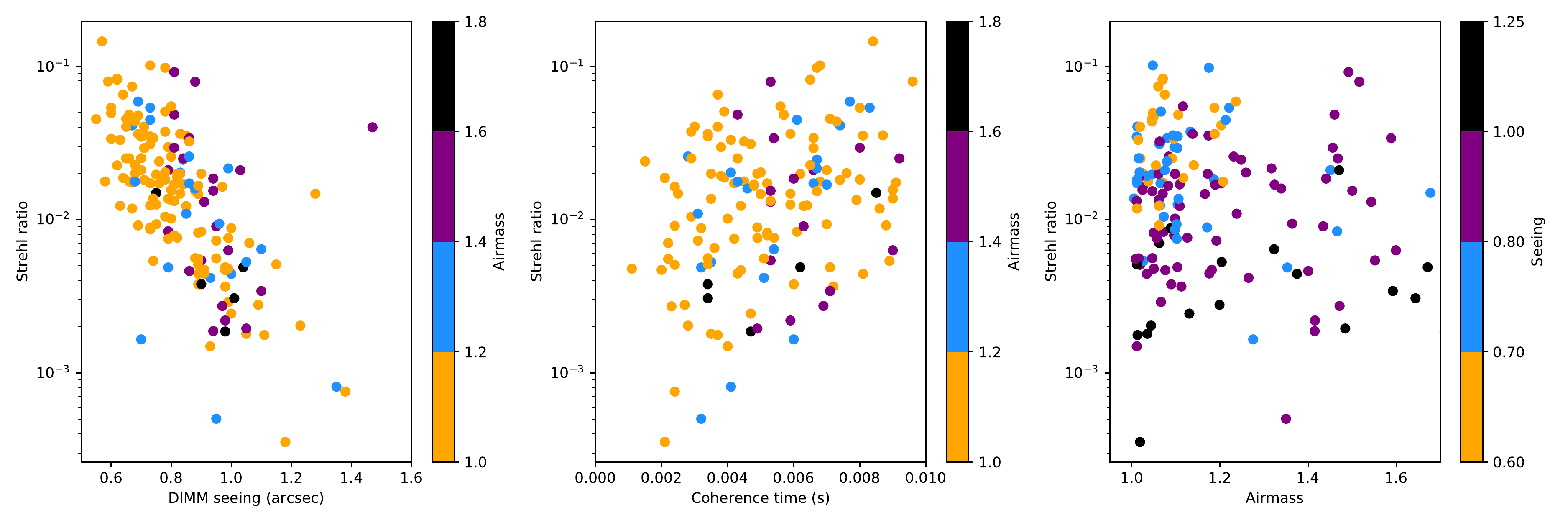}
    \caption{Top panel: PSF FWHM (measured from the {\tt maopppy} best fit parameters) as a function of ambient parameters (left: line of sight seeing, middle: coherence time, right: airmass). Bottom panel: same, but for the {\tt maoppy} Strehl ratio. The FWHM shows a weak dependence on seeing and coherence time, while the Strehl ratio shows a stronger degradation as the seeing increases. All values are shown at 650 nm.}
    \label{fig:strehl_ambipars_maoppy}
\end{figure}

\begin{figure}
    \centering
    \includegraphics[width=\textwidth]{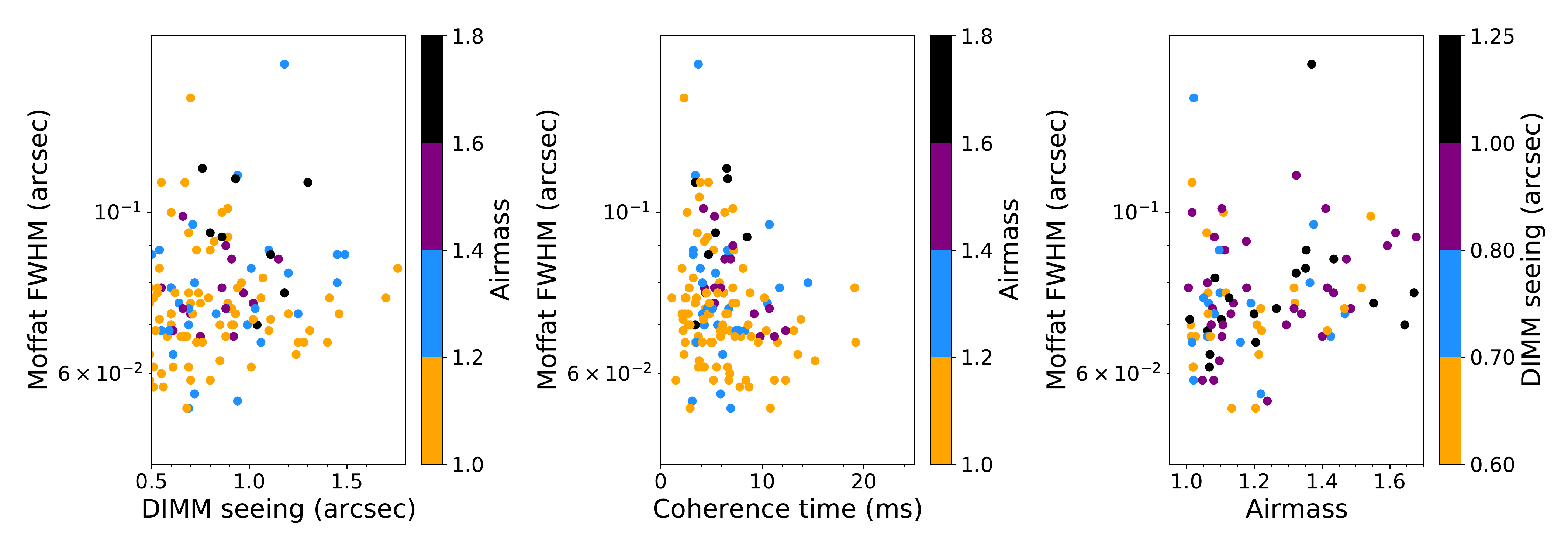}
    \centering
    \includegraphics[width=\textwidth]{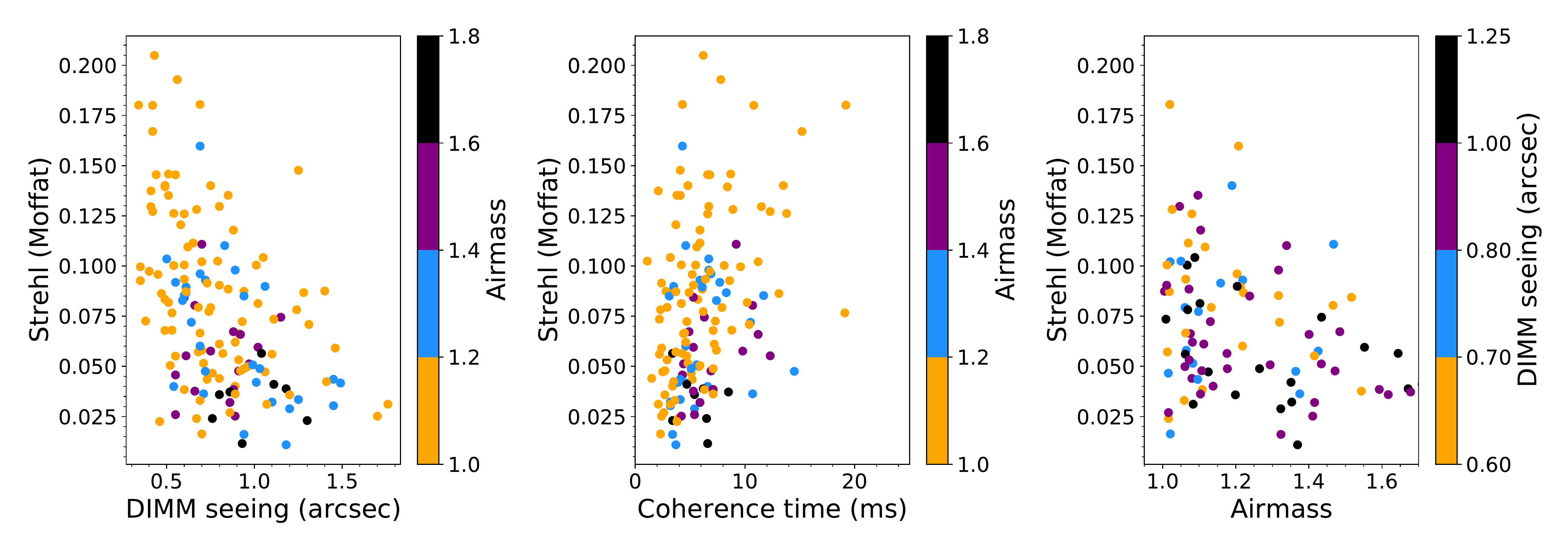}
    \caption{Top panel: PSF FWHM (measured from the Moffat best fit parameters) as a function of ambient parameters (left: line of sight seeing, middle: coherence time, right: airmass). Bottom panel: same, but for the Moffat Strehl ratio. The FWHM shows a weak dependence on seeing and coherence time, while the Strehl ratio shows a stronger degradation as the seeing increases. All values are shown at 682 nm.}
    \label{fig:strehl_ambipars_moffat}
\end{figure}

A different representation, which allows a more direct study of the dependence of the performance metrics on a given atmospheric variable, is shown in Figures \ref{fig:strehl_ambipars_maoppy} and \ref{fig:strehl_ambipars_moffat}. Here we plot the FWHM and Strehl ratio as a function of each individual environmental variable.
These plots indicate, for example, that the Strehl ratio has only a weak dependence on coherence time but a clear and strong dependence on the seeing conditions.

\begin{figure}
    \centering
    \includegraphics[width=0.75\textwidth]{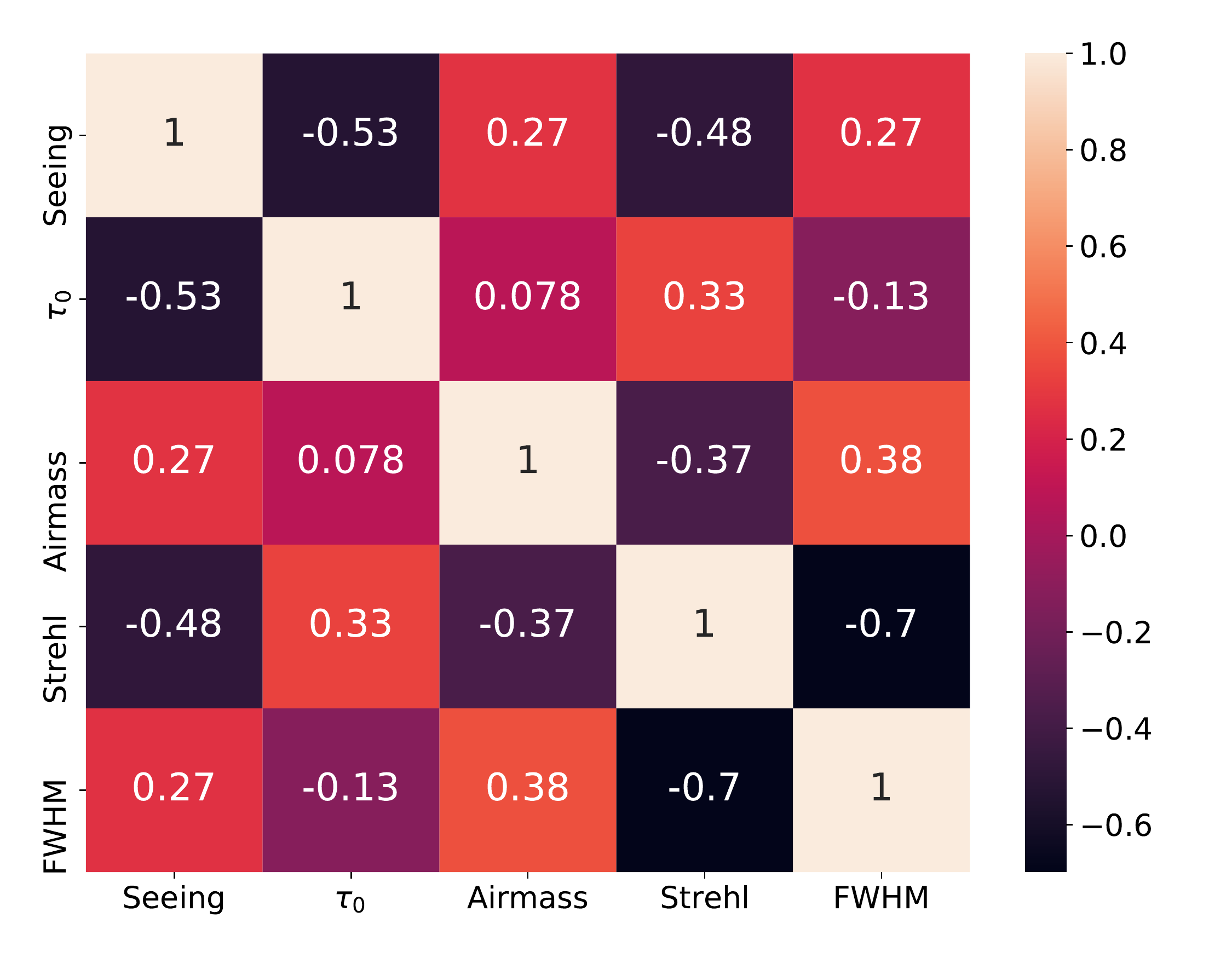}
    \caption{Spearman correlation matrix between environment variables (seeing, coherence time $\tau_0$ and airmass) and the performance variables (Strehl and FWHM of the Moffat model at 682 nm). }
    \label{fig:corrmat}
\end{figure}

More quantitatively, we present the Spearman correlation matrix between the environment variables and the Moffat variables (FWHM and Strehl) in Figure \ref{fig:corrmat} (we do not show the Pearson correlation matrix because some relationships are known to be non-linear). We start by noting that the seeing and coherence time correlate only weakly with airmass, as expected for values at zenith. The strongest correlations are observed for the Strehl ratio with seeing, airmass and coherence time, in that order. For the FWHM the correlations are similar, but generally weaker. In both cases, seeing and airmass are the dominant factors in determining the AO system performance. 

We attempt to quantify this further by performing a correlation analysis of the Strehl ratio (FWHM) with seeing and coherence time separately. From the left panel of Figure \ref{fig:strehl_ambipars_maoppy}, showing Strehl ratio as a function of DIMM seeing, there is a clear and robust correlation between these two parameters. The middle panel shows the Strehl ratio as a function of coherence time, which is significantly less pronounced. Finally, the right panel is shown to illustrate the effect of airmass. More quantitatively, we use a Bayesian approach to linear regression (i.e. of the functional form $y = a \times x + b$) implemented in the {\tt linmix} Python package \cite{Kelly07} to determine the correlation coefficients. We find the following correlations, where we quote posterior distribution median estimates, and the uncertainties are the standard deviations of the posteriors:
\begin{equation*}
\begin{array}{l}
  \rm   Strehl = (0.1\pm0.01) - (0.06\pm 0.01) \times DIMM\ seeing\\
  \rm FWHM = (0.074\pm0.005) + (0.012\pm0.007) \times DIMM\ seeing
  \end{array}
\end{equation*}
with a correlation coefficient $\hat{\rho}$ = --0.47$\pm$0.06 for the Strehl ratio and $\hat{\rho}$ = 0.16$\pm$0.08 for the FWHM (where $\hat{\rho}$=1 indicates a perfect correlation, and $\hat{\rho}$=0 indicates no correlation). Here the seeing is given in arcseconds.
For the coherence time (provided in ms), we find 
\begin{equation*}
\begin{array}{l}
  \rm   Strehl = (0.0032\pm0.006) - (0.0038\pm0.0009) \times \tau_0\\
  \rm   FWHM = (0.091\pm0.003) - (0.0011\pm0.0005) \times \tau_0
\end{array}
\end{equation*}
with correlation coefficients of $\hat{\rho}$ = 0.35 $\pm$0.07 for Strehl and $\hat{\rho}$ = -0.18 $\pm$0.08 for FWHM. 

We conclude that to first order, the effect of seeing on the NFM-AO performance (as measured through the Strehl ratio) significantly outweighs the degrading effect of poor coherence time.

A different method to quantify the relation between the environment variables and the performance variables, which can be useful for providing predictions given a set of atmospheric conditions, is multi-variate linear regression. The underlying assumption is that there is a linear relation between these variables. A complication for the current dataset is a significant anti-correlation between the seeing and coherence time, which may lead to a large variance in the correlation coefficients. This can be minimized by using ridge regression, but because the dependence of Strehl and FWHM on seeing and airmass are stronger than on coherence time, we ignore it for this work. 
The best fit multi-variate regression lines are given by:
\begin{equation*}
\begin{array}{l}
  \rm   FWHM = 0.04 + 0.0025 \times seeing + 0.032 \times airmass  -0.0007 \times \tau_0 \\
  \rm   Strehl = 0.19 -0.043 \times seeing - 0.08 \times airmass  + 0.0034\times \tau_0 \\
\end{array}
\end{equation*}
For typical values of seeing = 1.0 (0.5) arcsec, $\tau_0$ = 3.2 (7) ms and airmass = 1, we find FWHM = 72 (68) mas and a Strehl = 7 (10) per cent. This is roughly consistent with the values reported in  Table \ref{tab:performance_TC}.

We caution that the coefficient of determination (also called the r$^2$ statistic) for these multi-variate regression lines is relatively small (r$^2$ = 0.34). This indicates that either the assumption of a linear relation is flawed, or that there are additional parameters that introduce scatter into the relationship. Further investigation is required to better understand the predictive power of the simple analysis provided here. The main take-away point is that the system performance is determined mainly by the seeing and the airmass and, to a lesser extent, the coherence time.

\section{DISCUSSION AND SUMMARY}
\label{sec:summary}
We have presented an analysis of the system performance for the NFM-AO mode of the integral field unit MUSE at the VLT. Relying on two independent models to represent the resulting PSF, a dual Moffat profile as well as a more parametrized PSF implemented in the {\tt maoppy} Python package, we measure the FWHM and Strehl ratio of the PSF core (i.e. within the AO correction radius) for a set of STD observations. 
We investigate the behaviour of these two performance metrics with environment variables, including seeing, coherence time and airmass. 

We find that MUSE is more than capable of reaching its design Strehl ratio of 5\% at 650 nm. A Strehl ratio of $>5 \%$ can be delivered in all the currently allowed observing conditions, however airmass plays a key role in degrading performance when observations take place at airmass $>1.1$. Investigation of the FWHM as a function of turbulence category does not yield a strong dependence. The airmass does play an important role also for this metric. 
Investigation of different wavelength planes (at 498 nm and 912 nm) shows the expected degradation in the blue part of the spectrum, although the improvement towards redder wavelengths is clearly non linear for the FWHM. This may be partly understood by recalling that optical path aberrations can be similar to, or even larger than, the FWHM of the diffraction limited PSF.

While using STD observations is an excellent and flexible tool to characterize the performance of the NFM-AO system, there are several important caveats to the extrapolation of these results to science data. The first caveat is that we have only $\sim$40 datapoints following the upgrade to IRLOS+ in 2021 (partly due to an extended period of 6 months where one of the LGSUs was unavailable and NFM out of operations). These data are divided over the 5 relevant turbulence categories, and therefore we were forced to rely on data taken with the old IRLOS WFS to obtain a statistically relevant sample size. We also note that there are some parts of parameter space (e.g. observations at coherence times $\sim$2 ms, or at intermediate [1.2--1.5] airmass values) that are not well sampled by the current dataset, even when taking into account the observations obtained with the old IRLOS. As more data is accumulated with the upgraded system, we will be able to populate these parts of parameter space and obtain more accurate performance estimates in the future. Given the weak dependence of performance on coherence time, there may be prospects to expand the operational parameter space to lower coherence times.

The second caveat is that by definition, STD stars are bright enough to obtain high SNR data in relatively short exposure times. In contrast, science observations can and will target sources that are significantly fainter (up to $\sim$5 magnitudes) than STD stars. The current work does not, therefore, allow us to characterize the system performance at the faint end of the NGS magnitude distribution. This is particularly pressing because the upgraded IRLOS+ wavefront sensor has extended the limiting brightness by approximately 4~magnitudes, and this parameter space remains virtually unexplored both in the current work (due to the use of bright objects) and by previous commissioning activities (which showed almost flat performance figures down to the faintest magnitudes, albeit with very few data points). A dedicated observing campaign of a sample of sources spanning the faint end of the NGS brightness distribution would help to verify that the system performance does not suffer catastrophically in more challenging conditions.

This work will provide the first step towards further improving the MUSE exposure time calculator by allowing an empirical estimate of the expected image quality, SNR, Strehl ratio and PSF parameters to be implemented. This would provide a significant improvement over the current ETC, which uses a static PSF and therefore cannot account for differences in the observing conditions. In turn, this will allow science users to obtain the optimal exposure times as well as a more robust estimate of the required turbulence category for their future proposals. For example, it will inform users what the minimum resolving power of the NFM system is for disentangling close separation sources on the sky as a function of wavelength. An important development would be the construction of a software to determine the NFM PSF from telemetry and environmental variables, similar to the successful WFM-AO PSFr software\cite{2020A&A...635A.208F}. The results in this contribution show too large PSF variability to accomplish this in any useful manner. We hope that in the future, with more data and a full statistical study, this could be done.

In terms of science operations, a regular review of the system performance will facilitate the early discovery of potential problems and avoid the potential retrospective invalidation of acquired data due to unidentified problems. Because the current framework is scalable, it is also possible to include further telescope and AO telemetry that will facilitate the exploration of new parameter spaces. For example, there may be a wind direction dependence of the system performance that would be time consuming to investigate. By ingesting all DIMM telemetry, such investigations will become trivial and may yield a significantly improved knowledge of the system. This knowledge can in turn help to optimize AO systems for the next generation of extremely large telescopes.

\section*{Acknowledgments}
We thank the Paranal staff, and especially the TIOs (telescope Instrument Operators). Without their dedication and every night effort this work would not have been possible. We would also like to thank J\"oel Vernet and Dominika Wylezalek, whose scripts for the first MUSE NFM commissioning were the starting point for the development of those used for part of this work, and still at the heart of the dual-Moffat fits. We thank Ian J. Crossfield from the Astrobetter site for his python radial profile script. We would also like to acknowledge the use of the following python software packages: {\tt astropy}~\cite{astropy:2013, astropy:2018}, {\tt mstplotlib}\cite{Hunter:2007},  {\tt numpy}~\cite{harris2020array}, {\tt pandas}\cite{mckinney-proc-scipy-2010, reback2020pandas}, {\tt Photutils}\cite{larry_bradley_2020_4044744}, {\tt ScyPy}\cite{2020SciPy-NMeth}.

\bibliography{report} 
\bibliographystyle{spiebib} 

\end{document}